\documentclass{article}

\usepackage[table]{xcolor}
\usepackage{graphicx}
\usepackage{subcaption}
\usepackage{enumitem}
\usepackage{booktabs}
\usepackage{amssymb}
\usepackage[hidelinks]{hyperref}

\usepackage{dcase2026_techrep,amsmath,graphicx,url,times,booktabs, tabularx}

\title{
A multi-branch hierarchy-aware framework for heterogeneous audio classification}

\name{Beile Ning$^{1,\dagger}$\thanks{$\dagger$ These authors contributed equally to this work.},
      Jiayi Yu$^{1,\dagger}$,
      Zitong Wang$^{1,\dagger}$,
      Yufei Hu$^{1,\dagger}$,
      Wenjun Xu$^{1,\dagger}$}
\secondlinename{
      Yuanhang Qian$^{1}$,
      Zhongxin Bai$^{2}$,
      Gongping Huang$^{1,*}$\thanks{$^{*}$ Corresponding author.}
}

\address{$^1$ Wuhan University, Wuhan, China \\
         $^2$ Harbin Engineering University, Harbin, China \\
         \{ningbeile, gongpinghuang\}@whu.edu.cn
         }

\begin{document}

\ninept
\maketitle

\begin{sloppy}

\begin{abstract}

This technical report describes our system for Task 1 of the DCASE 2026 Challenge, which aims to classify heterogeneous audio recordings according to the Broad Sound Taxonomy (BST). The task requires both accurate second-level prediction and consistency with the top-level taxonomy. Our system is built on CLAP-based audio-text representations and is improved along three strategies: expanding the training set with a filtered subset of BSD35k, enhancing acoustic modeling with feature-specific branches, and refining predictions using hierarchy-aware classifiers and KNN-based post-processing. Among the acoustic features considered, the log-STFT branch provides the strongest single-model performance. With KNN-based post-processing, our best single system achieves a hierarchical F1 score (Hier. F1) of 80.84\% on the BSD10k-v1.2 set under the same evaluation protocol as the baseline. We further construct ensemble systems by combining models with complementary acoustic features and classification heads, achieving Hier. F1 of 81.25\% and 81.18\%, respectively.

\end{abstract}

\begin{keywords}
DCASE2026, CLAP, Heterogeneous Audio Classification, Broad Sound Taxonomy (BST)
\end{keywords}

\section{Introduction}
\label{sec:intro}
Heterogeneous audio classification aims to recognize sound events and acoustic scenes from real-world recordings with diverse content, recording conditions, and metadata quality. In DCASE 2026 Task 1, audio samples are annotated according to the Broad Sound Taxonomy (BST), where 23 second-level categories are grouped into 5 top-level classes~\cite{anastasopoulou2025general}. Since the official evaluation metric is based on the Hier. F1~\cite{kiritchenko2005functional}, a system should not only predict the correct second-level class, but also preserve consistency with the corresponding top-level category.

Our system uses CLAP audio-text representations~\cite{wu2023large} as the main semantic representation. To improve data diversity and reduce the impact of label noise, we construct an expanded training set, denoted as BSD-Grand, by incorporating a filtered subset of BSD35k~\cite{anastasopoulou_2026_19187100} into the BSD10k-v1.2 training data~\cite{anastasopoulou2024heterogeneous,anastasopoulou2025hierarchical}. The additional samples are selected through category-aware metadata cleaning, teacher-model filtering, and uploader-level constraints, which help reduce noisy annotations and uploader-specific bias. To complement the high-level semantic information captured by CLAP, we further introduce feature-specific acoustic branches based on MFCC, log-Mel spectrogram, and log-STFT features. Each acoustic feature is encoded by a corresponding branch and fused with the CLAP audio-text embeddings for audio classification. In addition, we investigate several prediction heads, including flat, global-classifier-based, and local-classifier-per-level-based heads, to better exploit the hierarchical relationship between the 5 top-level groups and the 23 second-level categories. Finally, we refine the model predictions using a KNN-based post-processing derived from the training embedding bank, and further incorporate this post-processing as soft supervision in a knowledge distillation framework.

Based on these components, we construct the following four submitted systems: \begin{itemize}
\item A single log-STFT-based model with KNN-based post-processing. This is our best single model and achieves a Hier. F1 of 80.84\%. 
\item An ensemble of KD-log-STFT, log-Mel, Flat, and LCL. This system is trained on the full training set.
\item An ensemble of KD-log-STFT, log-Mel, Flat, and LCL. This system is trained with 5-fold cross-validation and achieves the best overall performance, with a Hier. F1 of 81.25\%. 
\item An extended ensemble that further incorporates the MFCC branch and GC classification head. This system is trained with 5-fold cross-validation. It achieves a Hier. F1 of 81.18\%. 
\end{itemize} 
The experimental results show that dataset expansion, feature-specific acoustic modeling, and KNN-based prediction refinement consistently improve the CLAP-based baseline. The best ensemble system further benefits from the complementarity among different acoustic features and classification heads.

\section{Model Architecture and Training Strategy}
\subsection{Feature-Specific Acoustic Branch Framework}
\label{acoustic}

\begin{figure*}[h]
  \centering
  \includegraphics[width=0.88\textwidth]{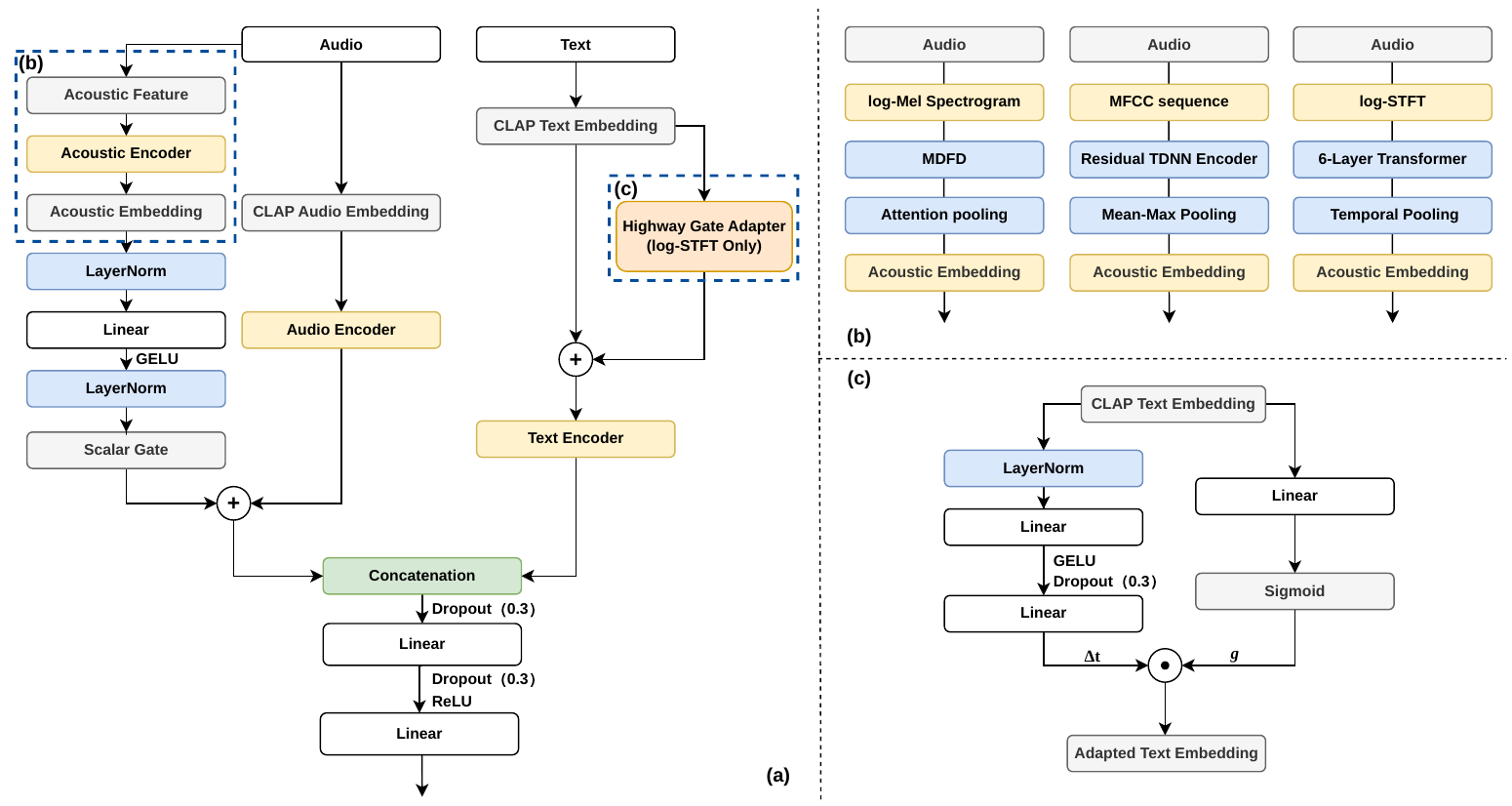}
  \caption{Overview of the proposed framework: (a) Overall architecture; (b) Feature-specific acoustic branches for log-Mel, MFCC, and log-STFT features; (c) Highway Gate Adapter.}
  \label{fig:base_structure}
\end{figure*}

As illustrated in Fig.~\ref{fig:base_structure}(a), we extend the original CLAP-based architecture by introducing two enhancement modules: a feature-specific acoustic branch (b) on the audio side and a Highway Gate Adapter (c) on the text side. The acoustic branches are designed to complement the high-level semantic representations extracted by the CLAP audio encoder with handcrafted acoustic features, whereas the Highway Gate Adapter performs task-specific adaptation of the CLAP text embedding through a gated residual connection. The enhanced audio representation and the adapted text representation are finally concatenated and fed into the classifier for 23 second-level classification.

The corresponding acoustic branches for different acoustic features are shown in Fig.~\ref{fig:base_structure}(b).
\begin{itemize}
\item \textbf{log-Mel branch}: log-Mel features are encoded by a Multi-Dilated Frequency Dynamic Convolution (MDFD) encoder followed by attention pooling.
\item \textbf{MFCC branch}: MFCC features are encoded by a residual Time Delay Neural Network (TDNN) with mean-max pooling.
\item \textbf{log-STFT branch}: log-STFT features are encoded by a 6-layer Transformer followed by temporal pooling.
\end{itemize}
Each branch encodes the corresponding acoustic feature into a compact acoustic embedding, which is then fused with the CLAP audio embedding through a gated residual fusion module.

The Highway Gate Adapter, shown in Fig.~\ref{fig:base_structure}(c), is applied only to the log-STFT branch. The adapter generates a task-specific residual representation, denoted as $\Delta\mathbf{t}$, from the original CLAP text embedding $\mathbf{t}$. A learnable highway gate~\cite{srivastava2015highway} is then introduced to adaptively control the contribution of $\Delta\mathbf{t}$ before it is combined with $\mathbf{t}$. The adapted text representation is computed as
\begin{equation}
\mathbf{t}'=\mathbf{t}+\mathbf{g}\odot\Delta\mathbf{t},
\end{equation}
where $\mathbf{g}$ denotes the learnable highway gate and $\odot$ denotes element-wise multiplication. This gated residual design enables task-specific refinement while preserving the pretrained semantic representation.

\subsection{Hierarchy-Aware Prediction}
\label{classifier}


Figure~\ref{fig:bst_frameworks} illustrates the classification heads used in our system. Inspired from~\cite{ding2023hierarchical},  We propose a baseline-based hierarchy-aware prediction framework. Specifically, we considered three variants: (a) a flat classification head, (b) a global classifier (GC) based head, and (c) a local classifier per level (LCL) based head. For all variants, we adopted the standard cross-entropy (CE) loss as the optimization objective:

\begin{itemize}
    \item \textbf{Flat classification head:} This approach directly projects the fused representations to the 23 second-level classes, completely ignoring the top-level taxonomy.Unlike the baseline, the flat classification head adds an extra linear layer compared to the baseline classification head. The loss function is simply the standard CE loss over the second-level nodes:
    \begin{equation}
        \mathcal{L}_\mathrm{Flat} = \mathcal{L}_\mathrm{second}.
    \end{equation}
    
    \item \textbf{GC classification head:} This variant adopts a multi-task learning architecture. The network outputs predictions for both the 5 top-level and 23 second-level categories simultaneously through parallel linear layers. The loss function is computed as the weighted sum of two standard CE losses:
    \begin{equation}
        \mathcal{L}_\mathrm{GC} = \mathcal{L}_\mathrm{second} + \lambda \cdot \mathcal{L}_\mathrm{top},
    \end{equation}
    where $\mathcal{L}_\mathrm{top}$ is the CE loss for the top-level categories and $\lambda$ is a balancing weight.
    
    \begin{figure*}[htbp]
  \centering
  \begin{subfigure}{0.32\textwidth}
    \centering
    \includegraphics[width=\linewidth]{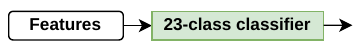}
    \caption{Flat classification head}
    \label{fig:a_flat}
  \end{subfigure}
  \hfill
  \begin{subfigure}{0.32\textwidth}
    \centering
    \includegraphics[width=\linewidth]{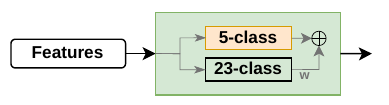}
    \caption{GC classification head}
    \label{fig:b_gc}
  \end{subfigure}
  \hfill
  \begin{subfigure}{0.32\textwidth}
    \centering
    \includegraphics[width=\linewidth]{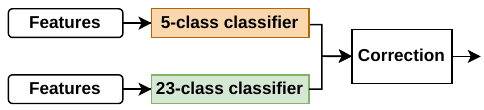}
    \caption{LCL classification head}
    \label{fig:c_lcl}
  \end{subfigure}

  \caption{The framework of the Hierarchical BST system}
  \label{fig:bst_frameworks}
\end{figure*}

    \item \textbf{LCL classification head:} The LCL head trains separate classifiers for different hierarchy levels. During training, it uses the exact same combined CE loss structure as the GC head:
    \begin{equation}
        \mathcal{L}_\mathrm{LCL} = \mathcal{L}_\mathrm{second} + \lambda \cdot \mathcal{L}_\mathrm{top}.
    \end{equation}
    However, during the inference phase, it employs a top-down correction mechanism (e.g., probability multiplication between the top-level and second-level outputs) to explicitly enforce hierarchical consistency and prevent top-level misclassifications from propagating downwards.
\end{itemize}


\subsection{KNN-Based Post-Processing}
\label{post}
We employ a KNN-based post-processing strategy to refine the final prediction by incorporating neighborhood information~\cite{kong2020knn,yang2022nearest}. 
A training embedding bank is first built from all training samples, where each retrieval embedding is formed by concatenating the feature representation from the trained log-STFT model with the original CLAP audio and text embeddings. 
For each evaluation sample, we extract its retrieval embedding in the same way and retrieve its $k$ nearest neighbors from the bank. 
The KNN-based neighbor prior is obtained by aggregating the neighbor labels with similarity-based weights:
\begin{equation}
p_{\mathrm{neighbor}}(c|z)
=
\sum_{i \in \mathcal{N}_k(z)}
w_i I(y_i=c),
\label{eq:neighbor_prior}
\end{equation}
where $\mathcal{N}_k(z)$ denotes the set of $k$ nearest neighbors of sample $z$, $y_i$ is the label of the $i$-th neighbor, $w_i$ is the normalized weight computed from temperature-scaled cosine similarity, and $I(y_i=c)$ is the class indicator.

The final prediction is computed as
\begin{equation}
p_{\mathrm{final}}
=
(1-\alpha)p_{\mathrm{model}}
+
\alpha p_{\mathrm{neighbor}},
\label{eq:knn_postprocess}
\end{equation}
where $p_{\mathrm{model}}$ is the original model prediction, $p_{\mathrm{neighbor}}$ is the KNN-based neighbor prior, and $\alpha$ controls its contribution.

To further integrate neighborhood information into training, we adopt a KNN-based distillation strategy. 
The log-STFT teacher constructs KNN-based target distributions for the training samples using the same embedding bank and neighbor aggregation process. 
The student model, which shares the same architecture as the teacher, is optimized using both the ground-truth labels and the KNN-based target distribution:
\begin{equation}
\mathcal{L}
=
\mathcal{L}_{\mathrm{CE}}
+
\lambda D_{\mathrm{KL}}
\left(
p_{\mathrm{neighbor}} \parallel p_{\mathrm{student}}
\right),
\label{eq:knn_distillation}
\end{equation}
where $p_{\mathrm{student}}$ denotes the student prediction, and $\lambda$ controls the weight of the distillation loss.

\subsection{BSD-Grand: Expanded Training Dataset Construction}
\label{dataset}
We constructed an expanded training set, referred to as BSD-Grand, by combining the full BSD10k-v1.2 dataset with a selected subset of BSD35k, while the validation and test sets keeping the original five-fold splits of BSD10k-v1.2 unchanged. The BSD35k subset was first restricted to the 23 target second classes. We then applied a category-aware metadata cleaning strategy, including additional filtering for uploader-specific templates and noisy descriptions. To reduce label noise, BSD35k samples were filtered using BSD10k-v1.2 five-fold teacher models under three different metadata views. Additionally an uploader-class cap of 200 samples was applied to reduce uploader-specific bias.

The final BSD-Grand training manifest contained 20,529 samples: 10,956 from BSD10k-v1.2 and 9,573 BSD35k selected samples. All training we had down was performed on this dataset.

\section{Experiments}

\subsection{Data Preprossing and Training Details}

All audio files were first resampled to 16 kHz and converted into fixed-duration 5-second clips. Audio clips shorter than 5 seconds were padded to the target length. Based on the processed waveform, we extracted three types of acoustic features: MFCC, log-Mel spectrogram, and log-STFT. More specifically, for MFCC features, we used 40 Mel filters, from which 13 cepstral coefficients were retained; for log-Mel features, we extracted 128 Mel frequency bins; for log-STFT features, we used an FFT size of 512, a window length of 400 samples, and a hop length of 160 samples. In addition, time masking~\cite{park2019specaugment} and random cropping~\cite{takahashi2019data} were applied to the time-frequency features during training to improve robustness to local temporal variations.

For text, since metadata quality varies across samples, we cleaned the text fields by removing non-acoustic metadata such as HTML tags and URLs from descriptions. Tags were split into individual terms and filtered to remove noisy or non-semantic tags. The cleaned text was then used to extract CLAP text embeddings.

For the single zero-byte audio file in the evaluation set, we used a silence fallback instead of discarding the sample. A zero waveform was used to extract both the spectral features and the CLAP audio embedding, ensuring that every evaluation item had a complete feature set.

In all experiments, we adopted the AdamW optimizer with an initial learning rate of $1\times10^{-3}$ and a weight decay of $1\times10^{-5}$. We trained each model for up to 100 epochs, using a batch size of 64, while models including feature-specific acoustic branches were trained with a batch size of 32. Early stopping was applied, which terminated training if the validation accuracy did not improve by more than 0.1\% over 15 consecutive epochs, while the patience window for models including feature-specific acoustic branches was shortened to 6 epochs. The loss function was cross-entropy with label smoothing of 0.05. We performed 5‑fold cross‑validation, where within each fold the training data was further split into training and validation sets using an 80/20 partition. All results were obtained with fixed random seeds to ensure reproducibility.

\subsection{Contributions of Feature-Specific Acoustic Branches and Hierarchy-Aware Prediction Heads}

Based on the architectures incorporating additional acoustic audio features introduced in Section~\ref{acoustic}, we obtain the corresponding training results as summarized in Table~\ref{acoustic_model}. In the table, the experimental configurations are defined as follows:
\begin{itemize}
    \item \textbf{Baseline}: Training on the BSD10k-v1.2 dataset.
    \item \textbf{+ BSD-Grand}: Training on the BSD-Grand dataset (Section~\ref{dataset}).
    \item All configurations below based on the BSD-Grand dataset use the corresponding acoustic feature branch structures described in Section~\ref{acoustic}:
        \begin{itemize} 
            \item[$\circ$] \textbf{+ BSD-Grand + log-Mel}: log-Mel branch.
            \item[$\circ$] \textbf{+ BSD-Grand + MFCC}: MFCC branch.
            \item[$\circ$] \textbf{+ BSD-Grand + log-STFT}: log-STFT branch.
            \item[$\circ$] \textbf{+ BSD-Grand + Post-log-STFT}: log-STFT branch with post-processing (Section~\ref{post}); best-performing single model.
            \item[$\circ$] \textbf{+ BSD-Grand + KD-log-STFT}: log-STFT branch with KNN-based knowledge distillation (Section~\ref{post}); used for model ensembling.
        \end{itemize}
\end{itemize}

\begin{table}[htbp]
  \centering
  \setlength{\tabcolsep}{2pt}
  \caption{Performance of models with different acoustic features (5-fold cross-validation)}
  \label{acoustic_model}
  \begin{tabular}{lrr}
    \toprule
    Config. & Hier. F1/\% & Hier. Accuracy/\% \\
    \midrule
    Baseline & 78.45 & 79.58 \\
    \hspace*{1em} + BSD-Grand & 79.64 & 80.61 \\
    \hspace*{1em} + BSD-Grand + log-Mel & 79.95 & 80.63 \\
    \hspace*{1em} + BSD-Grand + MFCC & 80.13 & 80.39 \\
    \hspace*{1em} + BSD-Grand + log-STFT & 80.54 & 81.12 \\
    \hspace*{1em} + BSD-Grand + Post-log-STFT & 80.84 & 81.39 \\
    \hspace*{1em} + BSD-Grand + KD-log-STFT & 80.62 & 81.20 \\
    \bottomrule
  \end{tabular}
\end{table}


Based on the architectures using different  classifiers introduced in Section~\ref{classifier}, we obtain the corresponding training results as summarized in Table~\ref{tab:classifier_model}. In the table, the experimental configurations are defined as follows:

\begin{itemize}
    \item \textbf{Baseline}: Training on the BSD10k-v1.2 dataset.
    \item \textbf{+BSD-Grand}: Training on the BSD-Grand dataset (Section~\ref{dataset}).
    \item All configurations below based on the BSD-Grand dataset use the corresponding classifiers described in Section~\ref{classifier}:
    \begin{itemize} 
        \item[$\circ$] \textbf{+ BSD-Grand + Flat}: Flat classification head.
        \item[$\circ$] \textbf{+ BSD-Grand + GC}: GC classification head.
        \item[$\circ$] \textbf{+ BSD-Grand + LCL}: LCL classification head.
    \end{itemize}
\end{itemize}

\begin{table}[htbp]
  \centering
  \setlength{\tabcolsep}{2pt}
  \caption{Performance comparison of different classifiers (5-fold cross-validation)}
  \label{tab:classifier_model}
  \begin{tabular}{lrr}
    \toprule
    Config. & Hier. F1/\% & Hier. Accuracy/\% \\
    \midrule
    Baseline & 78.45 & 79.58 \\
    \hspace*{1em} + BSD-Grand & 79.64 & 80.61 \\
    \hspace*{1em} + BSD-Grand + Flat & 80.57 & 81.53 \\
    \hspace*{1em} + BSD-Grand + GC & 80.01 & 81.02 \\
    \hspace*{1em} + BSD-Grand + LCL & 80.20 & 81.18 \\
    \bottomrule
  \end{tabular}
\end{table}

For brevity, in the following discussion we refer to each configuration by its last component. For example, ``+ BSD-Grand + KD-log-STFT'' in Table~\ref{acoustic_model} is abbreviated as ``KD-log-STFT'' and ``+ BSD-Grand + Flat'' in Table~\ref{tab:classifier_model} is abbreviated as ``Flat''.

\subsection{Submitted Systems}
\label{submission}
The performance of our submitted systems on the BSD10k-v1.2 dataset is summarized in Table ~\ref{tab:submission_bsd10k}. 
\begin{itemize}
  \item \textbf{System 1 (single model)}: Built on the log-STFT architecture with post-processing, this is the best-performing single model among all submissions.
  \item \textbf{System 2 (ensemble, full training)}: An ensemble of KD-log-STFT, log-Mel, Flat and LCL, trained once on the BSD-Grand dataset. The ensemble averages the logits with weights: 0.4 (KD-log-STFT), 0.2 (log-Mel), 0.3 (Flat), and 0.1 (LCL).
  \item \textbf{System 3 (ensemble, 5-fold cross-validation)}: Same architecture as System 2, but trained with 5-fold cross-validation. This variant achieves the highest Hier. F1 among all systems.
  \item  \textbf{System 4 (extended ensemble)}: Extends System 2 by additionally incorporating MFCC and GC. The logit averaging weights are: 1/11 (log-Mel), 1/11 (MFCC), 3/11 (KD-log-STFT), 2/11 (Flat), 1/11 (GC), and 3/11 (LCL).
\end{itemize}

\begin{table}[htbp]
  \centering
  \setlength{\tabcolsep}{2pt}
  \caption{Performance of submitted systems (5-fold cross-validation)}
  \label{tab:submission_bsd10k}
  \begin{tabular}{lcrr}
    \toprule
    Model & Ensemble & Hier. F1/\% & Hier. Accuracy/\% \\
    \midrule
    Baseline & $\quad$ & 78.45 & 79.58 \\
    System 1 & $\quad$ & 80.84 & 81.39 \\
    System 2 & $\checkmark$ & -- & -- \\
    System 3 & $\checkmark$ & 81.25 & 81.86 \\
    System 4 & $\checkmark$ & 81.18 & 81.79 \\
    \bottomrule   
  \end{tabular}
  
  \vspace{2pt}
  {\footnotesize Note: System 2 is not evaluated on the BSD10k‑v1.2 because its training consumes the entire dataset, leaving no held‑out test samples.}
\end{table}

\section{CONCLUSION}

This technical report describes our system for DCASE 2026 Task 1. We improve the CLAP-based baseline through BSD-Grand dataset expansion, multi-branch acoustic feature extraction (log-Mel, MFCC, log-STFT), hierarchy-aware classification heads, and KNN-based post-processing with knowledge distillation. Among single models, the log-STFT branch with post-processing achieves the best performance (80.84\% Hier. F1). Our best ensemble system (System 3) attains 81.25\% Hier. F1, significantly surpassing the baseline of 78.45\%. The results highlight the complementary benefits of acoustic feature diversification and ensemble learning for heterogeneous audio classification under the BST taxonomy.

\bibliographystyle{IEEEtran}

\end{sloppy}
\end{document}